\documentclass[11pt,a4paper]{article}
\pdfoutput=1

\usepackage{jcappub}

\usepackage{xcolor}

\usepackage{amsmath}
\usepackage{amsfonts,amssymb}
\usepackage{hyperref}
\usepackage{graphicx}
\usepackage{natbib}
\usepackage[utf8]{inputenc}

\begin{document}

\title{Stochastic collapse}

\author{Tays Miranda$^{1,2}$}
\emailAdd{tays.miranda@cosmo-ufes.org}

\author{Emmanuel Frion$^{2,3}$}
\emailAdd{emmanuel.frion@cosmo-ufes.org}

\author{David Wands$^{2}$}
\emailAdd{david.wands@port.ac.uk}

\affiliation{$^1$ Physics Department, Universidade Federal do Esp\'irito Santo, Avenida Fernando Ferrari 514, 29075-910 Vit\'oria, Esp\'irito Santo, Brazil,}
\affiliation{$^2$ Institute of Cosmology and Gravitation, University of Portsmouth, Dennis Sciama Building, Burnaby Road, Portsmouth, PO1 3FX, United Kingdom,}
\affiliation{$^3$ PPGCosmo, Universidade Federal do Esp\'irito Santo, Avenida Fernando Ferrari 514, 29075-910 Vit\'oria, Esp\'irito Santo, Brazil.}

\abstract{
We investigate cosmological models described by a scalar field with an exponential potential, and apply the stochastic formalism, which allows us to study how quantum field fluctuations give rise to stochastic noise. This modifies the classical dynamics of the scalar field at large scales, above a coarse-graining scale. In particular we explore how quantum field fluctuations perturb the equation of state on large scales which can lead to a quantum instability of the classical collapse solution below the Planck scale in the case of a pressureless collapse.


}

\maketitle

\section{Introduction}
\label{Introduction}
A common ansatz for the very early evolution of our Universe is cosmological inflation \cite{Starobinsky:1980te, Sato:1980yn, Guth:1980zm, Linde:1981mu,albrecht1982cosmology,linde1983chaotic,mukhanov1981quantum,mukhanov1982vacuum,starobinsky1982dynamics,guth1982fluctuations,hawking1982development,bardeen1983spontaneous}. This accelerated expansion in the very early universe is a well-tested paradigm since its predictions for the primordial power spectrum can be successfully compared with observations of the cosmic microwave background (CMB) \cite{Hinshaw:2012aka, Akrami:2018odb}. However, despite all the success of inflation in explaining large-scale observations, there are models seeking to give an alternative explanation of the initial conditions for the Big Bang and even the initial singularity. The most common such theories are called bouncing models, which include a universe before the Big Bang, often with a very long contracting phase, see \textit{e.g.} \cite{novello2008bouncing, cai2014exploring, battefeld2015critical,lilley2015bouncing, brandenberger2017bouncing, peter2008cosmology,Galkina:2019pir}, and possibly an inflationary phase after the bounce, see \textit{e.g.} \cite{biswas2012stable, piao2004suppressing, xia2014evidence, liu2013obtaining, qiu2015g, frion2019affine}. An early proposal in which a pre-Big Bang phase could set the initial conditions for a subsequent post-Big Bang phase was made by Gasperini and Veneziano \cite{Gasperini:1992em}. Since then, several models have been presented, see e.g. \cite{Khoury:2001wf, wands:2009cosmological}.

A scalar field with an exponential potential provides a simple model for an accelerated expansion and inflation in the early universe \cite{PhysRevD.32.1316, kitada1993cosmic, wands1993exponential} or dark energy at the present epoch \cite{Wetterich:1994bg, Amendola:1999qq}. But it can also drive a collapsing universe and this configuration is particularly interesting since due to their scale-invariant form, exponential potentials are simple to study analytically. A previous study with such a configuration in the case of bouncing models with the bounce due to quantum effects was done in \cite{bacalhau}. In order to explore this simplicity, we will focus our attention on models with a scalar field $\varphi$ and scalar potential
\begin{align}
V=V_0 \exp{\left(- \kappa \lambda \varphi \right)} \; ,
\end{align}
where $\kappa = \sqrt{8\pi G}$ and $\lambda$ is the slope of the potential. We can identify three scalar field collapse scenarios based on the form of the potential. In terms of energy density $\rho$ and pressure $P$ of the scalar field, we have 
\begin{itemize}
	\item Non-stiff/Matter/Pressureless collapse ($P < \rho$ with $V > 0$);
	\item Pre-Big Bang collapse ($P = \rho $ with $V = 0$);
	\item Ekpyrotic collapse ($P \gg \rho$ with $V < 0$);	
\end{itemize}	

Although the classical stability of collapsing models has already been studied in previous works  \cite{Heard:2002dr}, it is interesting to consider how the classical solutions behave in the presence of quantum fluctuations. To do this we will extend the stochastic formalism, previously introduced to study inflation, to collapse scenarios.

The stochastic formalism models the effect of quantum vacuum fluctuations by introducing a cut-off scale (the so-called coarse-graining scale) splitting the fluctuations into two parts: quantum vacuum modes (below the coarse-graining scale) and the long wavelength field which includes a stochastic noise. It was introduced in cosmology by Starobinsky \cite{Starobinsky:1986fx} to describe the effects of random vacuum fluctuations on inflationary dynamics \cite{nambu1988stochastic,nambu1989stochastic,kandrup1989stochastic,nakao1988stochastic,nambu1989stochastic2,mollerach1991stochastic,linde1994big,starobinsky1994equilibrium,Glavan:2017jye,Hardwick:2019uex,Saitou:2019jez,Tokuda:2018eqs,Hardwick:2018sck,Ando:2018qdb,Rigopoulos:2016oko,Nambu:2009zb,Tye:2008ef,Kunze:2006tu,Geshnizjani:2004tf,Matarrese:2003ye,Bellini:2001jm,Bellini:2000er,Bellini:1999zb,Calzetta:1999zr,Casini:1998wr,Ramsey:1997fz,Boyanovsky:1996ab,GarciaBellido:1994vz,Mijic:1994vv,Habib:1992ci,Mizutani:1991xh,Graziani:1988yd,Pattison:2017mbe,Pattison:2018bct}. More recently, this formalism was used in \cite{Grain:2017dqa} to show that slow-roll inflation is a stochastic attractor. 
The presence of a stochastic noise can affect the stability of a dynamical system in a non-trivial way. The aim of this paper is to investigate how these stochastic perturbations can modify the equation of state of the inflationary or collapse cosmology.

This paper is arranged as follows. In Sec.\ref{sec:2} we review the classical dynamics of a scalar field cosmology with an exponential potential, and we discuss the phase-space portrait for this theory as well as the stability of the fixed points which represent power-law expansion or collapse. In Sec.\ref{sec:3} we study linear perturbations about the background solutions and in particular the solutions of the perturbed field in a collapsing scenario. In Sec.\ref{linear noises} we describe the quantum fluctuations in the form of a stochastic noise on large scale and we study the deviations about the classical solution in phase-space, as well as the maximum lifetime of collapsing scenarios in the presence of stochastic fluctuations. We present our conclusions in Sec.\ref{conclusion}.

\section{Analysis of a scalar-tensor theory with exponential potential}\label{sec:2}

In this section we perform a qualitative analysis of a scalar-tensor cosmological model, for a scalar field $\varphi$ with an exponential potential. 

\subsection{Background cosmology}

We will consider a minimal coupling between gravity and a scalar field given by the action
\begin{align}
S = \int \sqrt{-g}\left[\frac{1}{2\kappa^{2}}R -\frac{1}{2} \partial^{\mu} \varphi \partial_{\mu} \varphi - V(\varphi)\right]d^{4}x\;.
\end{align}
%
Using the flat FLRW metric
\begin{align}
ds^2 = -dt^2+a^2(t)\delta_{ij} dx^{i} dx^{j} \;,
\end{align}
where $a$ is the scale factor, 
we obtain the equation of motion
\begin{align}
\ddot{\varphi} + 3 \frac{\dot{a}}{a} \dot{\varphi} + \frac{\partial V(\varphi)}{\partial\varphi} &= 0\;,\label{eq:KGback} 
\end{align}
subject to the Friedmann constraint
\begin{align}
3\left(\frac{\dot{a}}{a}\right)^{2} &= \kappa^{2}\left[\frac{1}{2}\dot{\varphi}^{2}+ V(\varphi)\right]\;,\label{eq:friedmann1}
\end{align}
where a dot stands for differentiation with respect to cosmic time. 

The energy density $\rho_{\varphi}$ and pressure $P_{\varphi}$ for the scalar field are given by
\begin{align}
\label{rhovarphi}
\rho_{\varphi} &= \frac{1}{2}\dot{\varphi}^{2} + V(\varphi)\;,\\
\label{Pvarphi}
P_{\varphi} &= \frac{1}{2}\dot{\varphi}^{2} - V(\varphi)\;.
\end{align}
Thus the parameter $w$ of the equation of state for $\varphi$ is given by
\begin{align}
w = \frac{P_{\varphi}}{\rho_\varphi} = \frac{\dot{\varphi}^{2} - 2V(\varphi)}{\dot{\varphi}^{2} + 2V(\varphi)}\;.\label{eq:eos1}
\end{align}

\subsection{Background phase-space}
In order to perform a qualitative analysis of the system described by (\ref{eq:KGback}) and (\ref{eq:friedmann1}), new variables can be introduced as
\begin{align}
\label{eq:dimensionless variables}
x = \frac{\kappa \dot{\varphi}}{\sqrt{6} H} \;, \quad y = \frac{\kappa \sqrt{\pm V}}{\sqrt{3} H},
\end{align}
where $H = \dot{a}/a$ is the Hubble rate 
and we use $\pm$ for positive and negative scalar potentials, $\pm V>0$. With these variables,
we can write the Friedmann constraint~(\ref{eq:friedmann1}) as
\begin{align}
x^2 \pm y^2 = 1\;,\label{eq:FriedmannConstraint}
\end{align}
and the equation of state~(\ref{eq:eos1}) becomes
\begin{align}
w = \frac{x^{2}\mp y^{2}}{x^{2}\pm y^{2}}\;.\label{eq:eos2}
\end{align} 

Then, we are able to rewrite the equation of motion
(\ref{eq:KGback}) in terms of the autonomous system
\begin{align}
x^{\prime} &= -3x(1-x^{2})\pm \lambda\sqrt{3/2}y^{2}\;,\label{eq:xback}\\
y^{\prime} &= xy(3x-\lambda\sqrt{3/2})\;,
\end{align} 
where a prime is a derivative with respect to the logarithm of the scalar factor and $N = \ln a$ counts the number of e-folds to the end of inflation or the collapse phase. 
We identify critical points of the system with fixed points where $x^{\prime}=0$ and $y^{\prime}=0$. 

There are two kinetic-dominated solutions 
\begin{align}
x_{a} = -1\ {\rm or}\ +1\;, \quad y_{a} = 0\;,\label{eq:critp1}
\end{align}
with equation of state~(\ref{eq:eos2}) $w_a=1$. These fixed points therefore correspond to solutions where $a\propto t^{1/3}$ in an expanding universe for $t>0$, or $a\propto (-t)^{1/3}$ for $t<0$ in a contracting universe.

There is also a potential-kinetic-scaling solution for $\pm(6-\lambda^{2})>0$ (i.e., a sufficiently flat positive potential, $\lambda^2<6$ for $V>0$, or a sufficiently steep negative potential, $\lambda^2>6$ for $V<0$) which is given by
\begin{align}
x_{b} = \frac{\lambda}{\sqrt{6}}\;, \quad y_{b}=\sqrt{\frac{\pm(6-\lambda^{2})}{6}}\;;\label{eq:critp2}
\end{align}

This scaling solution corresponds to a solution with constant equation of state~(\ref{eq:eos2})
\begin{align}
w_{b} &= \frac{\lambda^{2}}{3}-1\;,
\end{align}
and thus a power-law solution for the scale factor
\begin{align}
a(t) \propto |t|^{p}\;, \quad \varphi(t) = \sqrt{\frac{4}{3\kappa^{2}(1+w_b)}}\ln |t| + C\;,\label{eq:phisol}
\end{align}
where $C$ is an arbitrary constant of integration and
\begin{align}
\label{pval}
p = 
\frac{2}{\lambda^2} \,.
\end{align}

First-order perturbations around this critical point yield the linearised equation~\cite{Heard:2002dr}
\begin{align}
\label{classical x}
x^{\prime} = \frac{(\lambda^{2}-6)}{2} (x-x_{b})\;.
\end{align}
%
%
We see that in an expanding universe, $H>0$, the scaling solution (\ref{eq:critp2}) is stable for $\lambda^2<6$, corresponding to $p>1/3$ from (\ref{pval}). Thus the scaling solution is stable whenever it exists for a positive potential in an expanding universe, but it is never stable for a negative potential in an expanding universe. 
Conversely, in a collapsing universe, since $N$ decreases with cosmic time, $H<0$, the scaling solution is stable for $\lambda^2>6$, corresponding to $p<1/3$. Thus the scaling solution is stable whenever it exists for a negative potential in a collapsing universe, but it is never stable for a positive potential for $H<0$.

In summary:
\begin{itemize}
\item Expanding universe ($N\to+\infty$): \begin{itemize}
\item[$\diamond$] The scaling solution exists and is stable for a positive, flat potential $p>1/3$ (including inflation, $p>1$).
\item[$\diamond$] The scaling solution exists but is unstable for a negative, steep potential $p<1/3$.
\end{itemize}
\item Contracting universe ($N\to-\infty$): \begin{itemize}
\item[$\diamond$] The scaling solution exists and is stable for a negative steep potential $p<1/3$ (including ekpyrosis, $p\ll 1$).
\item[$\diamond$] The scaling solution exists but is unstable for a positive flat potential $p>1/3$ (including matter collapse, $p\simeq2/3$).
\end{itemize}
\end{itemize}

\section{Linear perturbations}
\label{sec:3}

\subsection{Scalar field and metric perturbations}
	
In order to investigate the effects of small scale quantum fluctuations on the stochastic evolution of the coarse-grained field above the Hubble scale
let us consider inhomogeneous scalar field perturbations, $\varphi = \varphi(t) + \delta\varphi(t,\vec{x})$, in a linearly-perturbed FLRW metric
\begin{align}
ds^2 = -(1+2A)dt^2+2a\partial_{i}B dx^{i}dt+a^2(t)\left[(1-2\psi)\delta_{ij}+2\partial_{ij}E+h_{ij}\right] dx^{i} dx^{j} \; ,
\end{align}
with $A$, $B$, $\psi$ and $E$ the scalar potentials and $h_{ij}$ the tensor perturbations. We obtain the wave equation for first-order scalar field perturbations
\begin{align}
\ddot{\delta\varphi} + 3H \dot{\delta\varphi} + \frac{k^{2}}{a^{2}}\delta\varphi + m^{2}\delta\varphi = - 2 V_{,\varphi} A
+ \dot{\varphi} \left[\dot{A} + \frac{k^{2}}{a^{2}}\left(a^{2}\dot{E} - a B\right)\right]\;,\label{eq:pertKG1}
\end{align}
where $k$ is the comoving wavenumber and $V_{,\varphi}$ is the derivative of $V$ with respect to the scalar field.
	
It has been shown in \cite{Mukhanov:1988jd,Sasaki:1986hm,wands:2009cosmological} that (\ref{eq:pertKG1}) in the spatially-flat gauge ($\psi=0$) can be written as
\begin{align}
&\ddot{\delta\varphi} + 3H\dot{\varphi} + \left[\frac{k^{2}}{a^{2}} + m^{2} - \frac{8\pi G}{a^{3}}
\frac{d}{dt}
\left(\frac{a^{3}\dot{\varphi}^{2}}{H}\right) \right]\delta\varphi = 0\;.
\end{align}
Then, we can introduce new variables $dt=ad\eta$, $v=a\delta\varphi$ and $z=a\dot{\varphi}/H$ to cast the above equation as a harmonic oscillator
\begin{align}
\label{eq:perteq}
\frac{d^2 v}{d\eta^2}+\left(k^2-\frac{1}{z}\frac{d^2  z}{d\eta^2}\right) v = 0\; .
\end{align}
Note that for a power-law cosmology, $z''/z=a''/a\propto (aH)^2$.
Hence the solutions for small (sub-Hubble) and  large (super-Hubble) scales are respectively
\begin{eqnarray}
\label{vacuumsoln}
\delta\varphi &\simeq& \frac{e^{-ikt/a}}{a\sqrt{2k}} \quad \textrm{for}\ k^{2}/a^2 \gg H^2
\,,\\
\delta\varphi &\simeq& \frac{C\dot{\varphi}}{H} + \frac{D\dot{\varphi}}{H} \int\frac{H^{2}}{a^{3}\dot{\varphi}^{2}}dt \quad \textrm{for}\ k^{2}/a^2\ll H^2
\,,
\end{eqnarray}
where we have chosen the quantum vacuum normalisation for the under-damped oscillations on sub-Hubble scales (\ref{vacuumsoln}).	

A characteristic feature of an inflating spacetime is that the comoving Hubble length decreases in an accelerating expansion with $\dot{a}>0$ and $\ddot{a}>0$. The same is true for the comoving Hubble length, $|H|^{-1}/a=1/|\dot{a}|$ in a decelerating, collapsing universe with $\dot{a}<0$ and $\ddot{a}<0$.
%
%
As a result quantum vacuum fluctuations, on sub-Hubble scales at early times (\ref{vacuumsoln}), lead to well-defined predictions for the power spectrum of perturbations on super-Hubble scales for potential-kinetic-scaling solutions with $\lambda^2<2$ (and hence $p>1$) in an expanding cosmology, or with $\lambda^2>2$ (and hence $p<1$) in a collapsing cosmology.
	
The characteristics of the inflation and collapse models for different values of $p$ are summarised in table \ref{comparison}.
\begin{table}[h]
\centering
\begin{tabular}{c c}
\large Power-law inflation \quad & \quad \large Decelerated collapse \vspace{0.1cm} \\ 
$H>0$ & $H<0$ \vspace{0.1cm}\\
$\dot{a}>0$, $\ddot{a}>0$ & $\dot{a}<0$, $\ddot{a}<0$ \vspace{0.1cm}\\ 
$p>1$ & $0<p<1$
\\
\end{tabular}
\caption{Comparing the quantities $H$, $\dot{a}$, $\ddot{a}$ and $p$ for power-law inflation and collapse. Although $\dot{a}$ is negative in the collapse case, its magnitude $\lvert\dot{a}\rvert$ is increasing. $p < 0$ is not allowed since this requires $\rho_\varphi+P_\varphi<0$, which is inconsistent with (\ref{rhovarphi}) and (\ref{Pvarphi}) for a canonical scalar field. }
\label{comparison}
\end{table}	\\

\subsection{Scalar field perturbations in a power-law collapse}\label{sec:ztochasticollapse}

Let us consider a collapsing universe with the scale factor being a power-law scaling solution with $a \propto (-t)^p$ and $t<0$ as in section \ref{sec:2}.
We can re-express the scale factor in terms of conformal time as
\begin{align}
\label{eq:scale factor}
a(\eta) \propto (-\eta)^{p/(1-p)}\;.
\end{align}
Using the relation for the Hubble rate in conformal time, $H=a^{\prime}/a^2$, we find that $a$ can also be expressed as
\begin{align}
a= \left( \frac{p}{1-p} \right) \frac{1}{H\eta} 
\;,
\end{align}
where $\eta<0$.
	
Since $\dot{\varphi}/H$ is constant in this case we have $z\propto a$, which allows us to rewrite (\ref{eq:perteq}) as a Bessel equation
\begin{align}
\label{harmonic oscillator}
\frac{d^2 v}{d\eta^2}+\left(k^2-\frac{\nu^2-1/4}{\eta^2}\right) v = 0 \;,
\end{align}
where
\begin{align}
\label{def:nu}
\nu = \frac{3}{2}+\frac{1}{p-1} = -\frac{3}{2}\left[\frac{1+(3p-2)}{1-(3p-2)}\right]\;.
\end{align}
Note that for power-law collapse with $p<1$ we have $\nu<3/2$\footnote{In Appendix \ref{AppMapping} we explicitly show the mapping between the quantities $p$, $\nu$, $\lambda^{2}$ and $w$.}.
	
The general solution for a given $k$ can be expressed as a linear combination of Hankel functions
\begin{align}
\label{Bessels}
v_k = \sqrt{|k \eta|} \left[V_{+}H_{\lvert\nu\rvert}^{(1)}(|k\eta|) + V_{-}H_{\lvert\nu\rvert}^{(2)}(|k\eta|)\right]\;.
\end{align}
We normalise the solution with the Bunch-Davies vacuum on small scales (early times), $v_k\to e^{-ik\eta}/\sqrt{2k}$ for $\eta \rightarrow -\infty$, so we set $V_{+} = 0$ and $V_{-} = \sqrt{\pi /4k}$. Equation~(\ref{Bessels}) then provides us with the corresponding solution on large scales (late times) for $k\eta \rightarrow 0$
\begin{align}
\label{eq:delta_phi}
\delta\varphi_{k} = \frac{i}{a}\sqrt{\frac{1}{4\pi k}}\frac{\Gamma(\lvert\nu\rvert)2^{\lvert\nu\rvert}}{\lvert k\eta\rvert^{\lvert\nu\rvert - 1/2}}\;.
\end{align}
As discussed in \cite{wands:2009cosmological}, this solution generates a scale invariant spectrum, $|\delta\varphi_k^2|\propto k^{-3}$, not only for slow-roll inflation ($P/ \rho \rightarrow -1$ and $\nu = 3/2$) but also for a pressureless collapse ($P/ \rho \rightarrow 0$ and $\nu = -3/2$).
	
Combining (\ref{eq:delta_phi}) and (\ref{eq:scale factor}), we see that in this large-scale limit the field perturbations are constant for $\nu >0$, since
\begin{align}
\delta\varphi_k \propto \frac{1}{a} (-\eta)^{\frac{1}{2}-|\nu |} \propto (-\eta)^{\nu - |\nu |} \;.
\end{align}
Conversely, for $\nu<0$ we see that the scalar field perturbations can grow rapidly on super-Hubble scales and diverge as $\eta\to0$.


\subsection{Perturbations in phase-space variables}


	

Introducing first order perturbations of the dimensionless phase-space variables (\ref{eq:dimensionless variables}), we obtain
\begin{align}
\label{eq:delta x total}
\delta x &= \frac{\kappa }{\sqrt{6} } \frac{1}{H} \left(\dot{\delta\varphi}-A\dot{\varphi}-\frac{\dot{\varphi}}{H}\delta H\right)\;,\\
\label{eq:perturbed y}
\delta y &= \frac{\kappa }{\sqrt{3}} \frac{\sqrt{V}}{H} \left(\frac{V_{,\varphi}}{2V}\delta\varphi-\frac{\delta H}{H}\right)\;,
\end{align}
where we are also including the metric perturbations $t\rightarrow(1+A)t$ and $H\rightarrow H+\delta H$ as described in \cite{Pattison:2019hef}. By perturbing the Friedmann equation, we obtain 
\begin{align}
\delta H =\frac{\kappa^{2}}{6H}(V_{,\varphi}\delta\varphi +\dot{\varphi}\dot{\delta\varphi}-\dot{\varphi}^{2}A)\;,
\end{align}
and since we are working in the spatially-flat gauge we can use the momentum constraint \cite{Mukhanov:1990me,Bassett:2005xm,wands:2009cosmological} to write the perturbed lapse function in terms of the scalar field perturbation as
\begin{align}
A = \frac{\kappa^{2}\dot{\varphi}\delta\varphi}{2H}\;.
\end{align}

At the critical point $x=x_{b}$ given by (\ref{eq:critp2}), the large-scale limit for the scalar field perturbations (\ref{eq:delta_phi}) then gives\footnote{The kinetic dominated solution analysis is briefly discussed in Appendix \ref{AppKinetic}, since it provides $\delta x = 0$ without regard to the solution $\delta\varphi$.}
\begin{align}
\delta x_k = \frac{i\kappa }{\sqrt{24\pi}}\left(1-\frac{\lambda^{2}}{6}\right)\left(\frac{2}{2\nu - 1}\right)^{2}\frac{\Gamma(\lvert\nu\rvert)2^{\lvert\nu\rvert}}{k^{\lvert\nu\rvert}}\left(\lvert\nu\rvert - \nu\right)\, H\, (-\eta)^{-\lvert\nu\rvert + 3/2}\;.
\end{align}
%
We can re-express $\delta x$ with respect to the Hubble rate at a given time
\begin{align}
\label{pivot}
H &= 
H_{\star} \exp{\left[\left(\frac{\nu-3/2}{\nu-1/2}\right) \left(N_{\star}-N\right)\right]} \; .
\end{align}
where $N_{\star}$ is evaluated at some initial time
and the conformal time can be expressed using (\ref{eq:scale factor}) as
\begin{align}
(-\eta)=(-\eta_{\star}) \exp{\left[\frac{1}{\nu-1/2}\left(N_{\star}-N\right)\right]} \;,
\end{align}
which gives
\begin{align}
\label{deltaxN}
\delta x_k = \frac{i\kappa }{\sqrt{24\pi}}\left(1-\frac{\lambda^{2}}{6}\right)\left(\frac{2}{2\nu - 1}\right)^{2+|\nu|}\frac{\Gamma(\lvert\nu\rvert)2^{\lvert\nu\rvert}}{\sigma^{\lvert\nu\rvert}}\left(\lvert\nu\rvert - \nu\right) H_{\star} (-\eta)^{3/2}_{\star} \exp{\left[\frac{\nu}{\nu-1/2} \left(N_{\star}-N\right)\right]}\;.
\end{align}
We see that for $\nu>0$, which includes power-law inflation ($\nu = 3/2$) and ekpyrotic collapse ($\nu = 1/2$), the scalar field perturbations at the leading-order on large scales leave the phase-space variable unchanged, $\delta x=0$.

Perturbing the Friedmann constraint \eqref{eq:FriedmannConstraint} requires $\delta y = -(x/y) \delta x$
and, as a consequence, we can write the perturbation of the equation of state \eqref{eq:eos2} as
\begin{align}
\delta w = 4x\delta x\;.
\end{align}
Hence for $\nu>0$ the scalar field perturbations at leading-order on large scales correspond to adiabatic perturbations which leave the equation of state unperturbed, $\delta w=0$. More generally, adiabatic perturbations on large scales correspond to local perturbations forwards or backwards in time along the background trajectory \cite{Wands:2000} which correspond to a fixed point in phase space. 
In appendix \ref{appnextorder} we consider the next-to-leading order scalar field perturbations on large-scales which give rise to a finite perturbation $\delta x\neq0$ on finite scales and finite time, $k\eta\neq0$ for $\nu>0$.

Conversely, for $\nu<0$ initial quantum field perturbations on sub-Hubble scales give rise to non-adiabatic perturbations on super-Hubble scales at late times, $\delta w\neq0$, which correspond to perturbations away from the fixed point in the dimensionless phase space.
We will now consider the effect of these quantum vacuum fluctuations evolving into the super-Hubble regime and giving rise to a stochastic diffusion in the phase space.

\section{Stochastic noise from quantum fluctuations}
\label{linear noises}
	
We will characterise the noise emerging from sub-Hubble quantum fluctuations evolving into the super-Hubble regime following the stochastic formalism described in \cite{Grain:2017dqa}. 
Classically, the time derivative of the field, $\dot{\varphi}$, can be related to its conjugate momentum, $\pi_{\varphi}\equiv \partial \mathcal{L}/\partial \dot{\varphi}$, by \cite{Grain:2017dqa}
\begin{align}
\dot{\varphi} = \frac{1}{a(t)^{3}}\pi_{\varphi}\;,\label{eq:phipi}
\end{align}
and the evolution for $\pi_{\varphi}$ is given by
\begin{align}
\dot{\pi}_{\varphi} = - a(t)^{3} V_{,\varphi}
\;,
\end{align}
where ``,$\varphi$" represents the derivative with respect to $\varphi$.

In the stochastic approach, one splits the scalar field and its momentum into a long-wavelength part and small-wavelength part as
 \begin{align}
 \varphi = \overline{\varphi} + \varphi_{Q}\,, \qquad \pi = \overline{\pi} + \pi_{Q}\;,
 \end{align}
 where the subscript ``$Q$" describes the small-wavelength quantum part. 
We introduce a time-dependent comoving cut-off scale (the so-called coarse-graining scale)
\begin{align}
\label{cgscale}
k_{\sigma} = \sigma a H \;,
\end{align}
below which small (sub-Hubble) wavelengths are integrated out, thus deriving an effective theory for the long wavelength part, with $\sigma$ the ratio between the Hubble radius and the cut-off wavelength. Note that we require the coarse-graining scale to be larger than the Hubble scale, hence $\sigma <1$, to neglect gradient terms \cite{Pattison:2019hef}.

In an inflationary expanding cosmology ($p>1$ and $t>0$) or a decelerated collapsing cosmology ($0<p<1$ and $t<0$) 
this leads to stochastic noise associated with the small-wavelength quantum fluctuations modes crossing the coarse-graining scale into the long-wavelength field at each time step, $d\tau$, is described by the two-point correlation matrix $\Xi(\vec{x}_1,\tau_1;\vec{x}_2,\tau_2)$
\begin{align}
\Xi_{f,g}(\vec{x}_1,\tau_1;\vec{x}_2,\tau_2) = \left\langle 0 | \xi_f (\vec{x}_1,\tau_1) \xi_g (\vec{x}_2,\tau_2) |0 \right\rangle\;,
\end{align}
with $f$, $g$ and $\xi_f$, $\xi_g$ being shorthand notation for the field or its momentum and their respective noises. In the following we will choose the time variable, $\tau$, to be the number of e-folds, $N$. Then, we can rewrite these entries in terms of the power spectrum (for white noise) as
\begin{align}
\Xi_{f,g}(N) &= \frac{1}{6\pi^{2}}\frac{dk^{3}_{\sigma}(N)}{dN}f_{k}\left(N\right)g_{k}^{*}\left(N\right) \label{eq:Xifg} 
\;.
\end{align}

\subsection{Quantum diffusion in phase-space}	
	
	
Near the critical point $x=x_b$ given by (\ref{eq:critp2}), the stochastic version of (\ref{classical x}) is\footnote{For more informations on stochastic differential equations, see e.g. \cite{oksendal2013stochastic}.}
\begin{align}
\frac{d(\bar{x}-x_b)}{dN} &= m (\bar{x}-x_b) +\hat{\xi}_{x},
\end{align}
where the eigenvalue $m=(\lambda^2-6)/2$, whose solution is given by considering an It\^o process
\begin{align}
\label{stochastic solution x}
\bar{x}(N)-x_c &= e^{m\left(N-N_{\star}\right)} \left(\bar{x}(N_{\star})-x_c\right) + \int_{N_{\star}}^{N} e^{m\left(N-S\right)} \hat{\xi}_{x} dS.
\end{align}
	
We define the variance associated with the coarse-grained field $\bar{x}$ as
\begin{align}
\sigma_x^2 := \left\langle \left(\bar{x}(N)-x_c\right)^2 \right\rangle \;;
\end{align}
whose evolution equation is given by
\begin{align}
\label{variancedifeq}
\frac{d\sigma_x^2}{dN} = 2 m \sigma_x^2+2 \left\langle \hat{\xi}_x \left(\bar{x}-x_c\right) \right\rangle .
\end{align}
The solution can be split into a classical part and a quantum part, given by\footnote{We show the explicit calculation in Appendix \ref{App2}.}
\begin{align}
\label{variance}
\sigma_x^2 (N) &= \sigma_{x,cl}^2 (N)+\sigma_{x,qu}^2 (N)\nonumber\\
&=\sigma_x^2(N_{\star}) e^{2m\left(N-N_{\star}\right)} + \int_{N_{\star}}^{N}  dS \, e^{2m(N-S)} \Xi_{x,x}(S)  \;,
\end{align}
where the two-point correlation matrix $\Xi_{x,x}(S)$ is defined in Eq.~(\ref{eq:Xifg}), using the notation of \cite{Grain:2017dqa}. The classical part, $\sigma_{x,cl}^2 (N)$, is given by the variance at some initial time times an exponential function of the number of e-folds. The quantum part, $\sigma_{x,qu}^2 (N)$, is the accumulated noise between the initial time and a later time. 

We find the two-point correlation function for the perturbations in the dimensionless phase-space variable, $x$, by applying \eqref{eq:Xifg} to $\delta x$ given in \eqref{deltaxN} as
\begin{align}
	\label{eq:Xixx}
	\Xi_{x,x}(N) &= \frac{1}{2\pi^{2}}\sigma^{3} \left(\nu -\frac{1}{2}\right)^2  \frac{1}{(-\eta_{\star})^3} \exp{\left[\frac{-3}{\nu-1/2} \left(N_{\star}-N\right)\right]} |\delta x|^2 \nonumber \\
&= g(\nu,\sigma) \kappa^{2} H^{2}(N)
\;,
\end{align}
with
\begin{align}
g(\nu,\sigma) := \frac{\Gamma^{2}(\lvert\nu\rvert)\nu^{2}2^{2\lvert\nu\rvert +4}}{(12\pi)^{3}\sigma^{2\lvert\nu\rvert - 3}} \left(\frac{2}{2\nu - 1}\right)^{2\lvert\nu\rvert + 4}(\lvert\nu\rvert -\nu)^{2}\;.
\end{align}
	
	
As previously noted, for $\nu>0$ the classical trajectory for $x$ remains preserved by the leading order perturbations in the scalar field since $\delta x = 0$ on large scales ($k\eta\to0$) \footnote{In Appendix \ref{appnextorder}, we take into account the next-to-leading order field contribution to compute $\Xi_{x,x}(N)$ and show that even in this configuration quantum diffusion should not take the system away from the fixed point. We show in particular that this is the case for quasi-de Sitter inflation.}, but the same is not true for $\nu$ negative since in this case $\delta x \neq 0$. We derive in Appendix \ref{appendixc} an alternative way to find this result using perturbations of the field and momentum. 


Inserting our result for the correlation function (\ref{eq:Xixx}) in the quantum part of (\ref{variance}), we find
\begin{align}
\label{variance final}
\sigma_{x,qu}^2 (N)	&= g(\nu,\sigma) \kappa^{2} H^{2}_{\star} \exp{\left[\frac{3-2\nu}{\nu-1/2} \left(-N_{\star}\right) \right]} e^{2mN}\int_{N_{\star}}^{N}  dS \, e^{-2mS}  \exp{\left[\frac{3-2\nu}{\nu-1/2} S \right]} \; .
\end{align}
Re-expressing the eigenvalue $m$ in terms of the index, $m=-2\nu/(\nu-1/2)$, the solution of (\ref{variance final}) is then
\begin{align}
\label{variance hn}
\sigma_{x,qu}^2 (N)
&= \tilde{g}(\nu,\sigma) \kappa^{2} H^{2}(N) \left\lbrace 1-\exp{\left[\frac{3+2\nu}{\nu-1/2}\left(N_{\star}-N\right)\right]}\right\rbrace \;,
\end{align}
where
\begin{align}
\tilde{g}(\nu,\sigma) &= \left(\frac{\nu-1/2}{3+2\nu}\right) g(\nu,\sigma)\nonumber\\
&=\frac{\Gamma^{2}(\lvert\nu\rvert)2^{2\lvert\nu\rvert +4}}{(12\pi)^{3}\sigma^{2\lvert\nu\rvert - 3}} \left(\frac{\nu^{2}}{3+2\nu}\right)\left(\frac{2}{2\nu - 1}\right)^{2\lvert\nu\rvert + 3}(\lvert\nu\rvert -\nu)^{2}\;.
\end{align}
Equation~(\ref{variance final}) is given in terms of the Hubble scale at a fixed time, $H_\star$, while we have used \eqref{pivot} to give the variance (\ref{variance hn}) in terms of the time-dependent Hubble scale, $H(N)$.

We can compare the growth rate of the classical and quantum perturbations by comparing the time dependence from the two parts in \eqref{variance}. We note first that the time dependence of the classical term goes as
\begin{align}
	\sigma_{x,cl} \propto \exp\left[\frac{4\nu }{\nu-1/2} \left(N_{\star}-N\right)\right]\;.
\end{align}
From \eqref{variance hn}, we see the time dependence of the quantum term behaves as
\begin{align}
 \sigma_{x,qu} \propto \exp\left[\frac{4\nu }{\nu-1/2} \left(N_{\star}-N\right)\right] \left\lbrace\exp\left[-\frac{3+2\nu }{\nu-1/2} \left(N_{\star}-N\right)\right] -1\right\rbrace \;.
\end{align}
Remember $N_{\star}-N$ grows with time ($N$ decreases) in an expanding universe. Thus, the quantum variance decays with time if we have
\begin{align}
\frac{3+2\nu}{\nu-1/2} >0 \;.\label{conditionfornu}
\end{align}
This is the case if either $\nu>1/2$ or $\nu<-3/2$. However, a positive $\nu$ will cancel the leading order quantum diffusion, so we will consider only the second case, $\nu<-3/2$, in the following analysis. Thus, the classical perturbations grow faster than the quantum noise if $\nu<-3/2$, and the quantum noise grows faster if $\nu>-3/2$.

Also, the condition \eqref{conditionfornu} provides a shift in the spectrum, when compared to the case $\nu = -3/2$, since the scalar spectral index can be written in terms as
 \cite{Zeldovich:1983cr,Wands:1998yp,Finelli:2001sr,Peter:2006hx,Peter:2008qz,Guimaraes:2019sqf}
\begin{align}
n_{\rm s} &= 1+ \frac{12 w}{1+3w} =  1+\frac{4(2\nu+3)}{3}\;,
\end{align}
and it is clear to see that when $\nu = -3/2-\epsilon$, where $\epsilon$ is a small positive parameter, $w < 0$ and the spectrum becomes red, \textit{i.e.}, $n_{\rm s} < 1$. 

To understand the behaviour around $\nu \approx -3/2$ we will consider $\nu=-3/2-\epsilon$ which for $|\epsilon(N_\star-N)|\ll1$ leads to
\begin{align}
\label{variance star pressure}
\sigma_{x,qu}^2 (N) 
&= \frac{3}{128 \pi} \frac{1}{\sigma^{2\epsilon}} \frac{H^{2}(N)}{M_{pl}^2}    \left(N_{\star}-N\right) \;,
\end{align}
where we have used $\kappa^{2} =8 \pi/M_{pl}^2$ with $M_{pl}$  the Planck mass, and we recall that $\sigma$ is the coarse-graining scale. The diffusion thus has the form of a random walk with $N_\star-N$ steps of equal, but growing, length $\propto |H(N)|$.

We see that \eqref{variance star pressure} depends weakly on the coarse-graining scale for 
$\nu \approx -3/2$, 
and becomes independent of $\sigma$
in the limit $\nu=-3/2$, where $\epsilon \rightarrow 0$.
This is not surprising since we know that quantum fluctuations in a pressureless collapse give rise to a scale-invariant spectrum of perturbations \cite{Wands:1998yp,Finelli:2001sr,Allen:2004vz}.
	
	
\subsection{Maximum lifetime of the collapsing phase}

We can now examine when the variance becomes large, i.e., when $\sigma_{x,qu}\approx1$, so that the quantum diffusion due to the stochastic noise results in a significant deviation from the critical point.

\subsubsection{Radiation-dominated collapse}

Consider first the case of a potential-kinetic-scaling collapse with $\lambda=2$, giving rise to an equation of state $w=1/3$, analogous to a radiation-dominated cosmology, and index $\nu=-1/2$. The variance \eqref{variance hn} in this case becomes
\begin{align}
\sigma_{x,qu}^2(N) &= \frac{\sigma^2}{54 \pi} \frac{H_{\star}^2}{M_{pl}^2} \left\{\exp{\left[4\left(N_{\star}-N\right)\right]}-\exp{\left[2\left(N_{\star}-N\right)\right]}\right\} \nonumber \\
&\approx \frac{\sigma^2}{54 \pi} \frac{H^2(N)}{M_{pl}^2} \;,
\end{align}
where to get the second line, we have neglected the second exponential term since the first one will grow much quicker. For $\sigma^2_{x,qu}(N_{\rm end})=1$, we get the straightforward result 
\begin{align}
|H_{\rm end}| \approx \frac{13}{\sigma} M_{pl} \; .
\end{align}

We conclude that a radiation-dominated collapsing phase cannot escape the fixed point due to quantum diffusion until it approaches the Planck scale. Indeed, since we require $\sigma <1$, we see that a deviation from the classical fixed point $x=x_b$ due to quantum diffusion would require the Hubble scale to become greater than the Planck scale. In practice as soon as the Hubble scale approaches the Planck scale our semi-classical analysis breaks down.

\subsubsection{Pressureless collapse}
	
\begin{figure}
\centering
\includegraphics[scale=0.3]{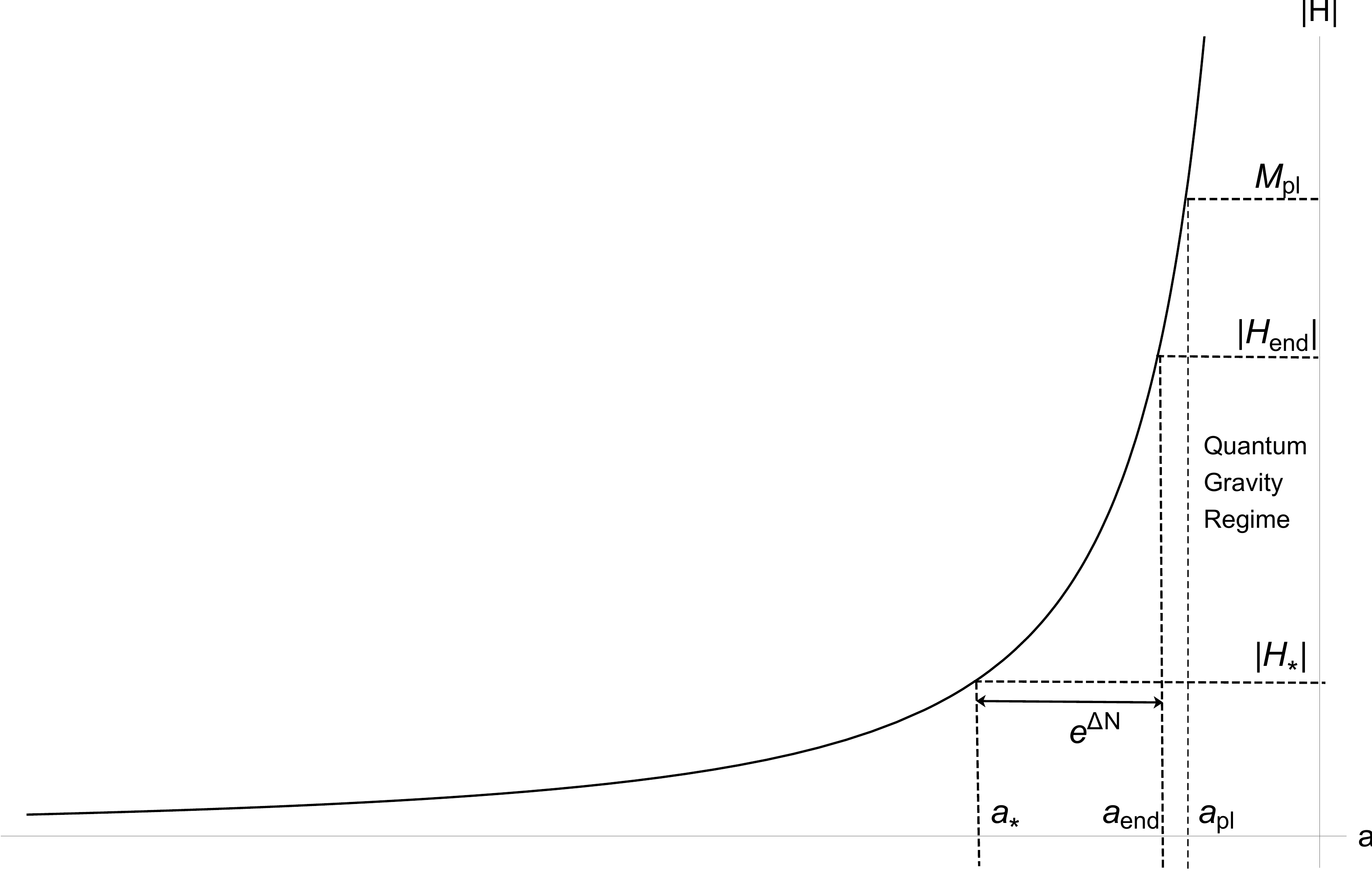}
\caption{Evolution of the Hubble rate, $\lvert H\rvert$, in a pressureless collapse. For quantum diffusion to lead to a deviation from the classical fixed point before the Hubble rate reaches the Planck scale, $|H_{\rm end}|<M_{\rm Pl}$, requires a very low initial energy scale, $|H_*|\ll |H_{\rm end}|$.
}
\label{scale}
\end{figure}

For the case of a pressureless collapse, $\nu = -3/2$, we find $\sigma_{x,qu}(N_{\rm end})=1$ when
\begin{align}
\label{lifetime}
|H_{\rm end}| = \sqrt{\frac{128 \pi}{3 (N_{\star}-N_{\rm end})}} M_{pl} 
\end{align}
Thus, for pressureless collapse case, quantum diffusion gives a time, $t_{\rm end}$, at which stochastic trajectories leave the classical fixed point before we reach the Planck scale, $|H_{\rm end}|<M_{pl}$, if the number of e-folds during the collapse is greater than 134. 
We show a simple example in Fig.\eqref{scale}. 

In terms of the initial Hubble rate, using \eqref{pivot} for the Hubble rate $H(N)$ for $\nu=-3/2$, we have
\begin{align}
(N_{\star}-N_{\rm end}) \exp{\left[3(N_{\star}-N_{\rm end})\right]} =\frac{128 \pi}{3} \frac{M_{pl}^2}{ H_{\star}^2} \;,
\end{align}
from which we get an approximate number of e-folds during the collapse phase
\begin{align}
N_{\star}-N_{\rm end} \approx \frac{2}{3} \ln{\left(\sqrt{\frac{128 \pi}{3}}  \frac{M_{pl}}{ |H_{\star}|}\right) } \;.
\end{align}
Conversely we can obtain an expression for the Hubble rate at the end of the pressureless collapse starting from an initial Hubble rate $H_{\star}$ given by
\begin{align}
|H_{\rm end}| \approx \sqrt{\frac{64 \pi}{\ln{\left(\sqrt{\frac{128 \pi}{3}}  \frac{M_{pl}}{ |H_{\star}|}\right) }}} M_{pl} \;.
\end{align}
	
Knowing how the comoving Hubble length behaves in terms of time during pressureless collapse, we can estimate a lower limit on the number of e-folds required during pressureless collapse to solve the horizon and smoothness problems of the hot big bang:
\begin{align}
\frac{k_{\rm end}}{k_\star} = \frac{a_{\rm end}H_{\rm end}}{a_\star H_\star} = \left( \frac{t_\star}{t_{\rm end}} \right)^{1/3} = e^{(N_\star-N_{\rm end})/2} > e^{70}\;,
\end{align}
where we are considering $70$ as a ratio between the Hubble length over Planck scale compared to horizon size today. This is a similar number for inflation to solve the flatness and  horizon  problems  of  Big  Bang  cosmology. Then, we would need 
\begin{align}
N_\star - N_{\rm end} > 140\;,
\end{align}	
which is remarkably close to the estimate $N_\star-N_{\rm end}>134$ that follows from requiring $|H_{\rm end}|<M_{pl}$ (\ref{lifetime}).


\section{Conclusion}
\label{conclusion}
	
The origin of the primordial density perturbations which lead to the large-scale structures we observe in the Universe is still a topic of debate. While a lot of attention has been given to quantum fluctuations in an inflationary expansion phase, another interesting possibility is that of vacuum fluctuations in a collapse phase preceding the present cosmological expansion. Perturbations of a self-interacting scalar field can be a useful starting point to study collapsing phases since in the case of linear perturbations, we can have a duality between the expanding and contracting solutions, as shown in \cite{Wands:1998yp}.

We have presented in this work an analysis of the stability of a power-law solution for a scalar field with an exponential potential, both classically and in the presence of quantum perturbations which can give rise to a stochastic noise on super-Hubble scales, either during an accelerated expansion or a decelerated collapse. We considered three possible cosmological scenarios: power-law inflation, a pressureless collapse 
and an ekpyrotic collapse. It is well known that inflationary and ekpyrotic models are classically stable under perturbations, while the matter collapse is unstable. We have shown in this work that in inflationary and ekpyrotic models quantum perturbations at leading-order on large (super-Hubble) scales are adiabatic, and do not drive these models away from their fixed point solution in phase-space. This means that the equation of state is unperturbed, $\delta w=0$, and one does not modify the classical behaviour of such models by adding quantum noise. In fact, all models with a positive index $\nu$ (the index of the Bessel function governing the evolution of the Mukhanov-Sasaki variable, see Eq.~(\ref{def:nu})) have a vanishing quantum contribution to the equation of state at leading order on super-Hubble scales. As a consequence, only models with a negative index, $\nu<0$, including the pressureless collapse case, diffuse away from the classical fixed point due to quantum noise. The perturbation of the equation of state
is summarised in Table \eqref{summary}.
	
\begin{table}[h]
\centering
\begin{tabular}{c|c}
\large Inflation or Ekpyrotic collapse  \quad & \quad \large Pressureless collapse  \\ 
\hline
$\nu > 0 \rightarrow$ adiabatic noise & $\nu < 0 \rightarrow$ non-adiabatic noise \\ 
$\delta w = 0$  & $\delta w\neq 0 $	\\
\end{tabular}
\caption{Table comparing the behaviour of $\delta w$ for three cases in the super-Hubble limit $\sigma \rightarrow 0$: de Sitter inflation ($\nu = 3/2$), ekpyrotic collapse ($\nu = 1/2$) and pressureless collapse ($\nu = -3/2$).}
\label{summary}
\end{table}

We then considered the maximum lifetime of the classical fixed point in the presence of quantum noise in a collapse phase with $\nu<0$. In the general case, we find that the collapse models are stable against quantum diffusion in the semi-classical theory since the variance with respect to the classical fixed point on super-Hubble scales is proportional to the Hubble rate and remains small while the Hubble rate remains below the Planck scale. 

However, for the particular case $\nu=-3/2$, corresponding to a pressureless collapse, the perturbations on super-Hubble scales are known to be scale-invariant; in this case we found the quantum diffusion leads to a random walk away from the fixed point. If we start the classical collapse from very low energy scales arbitrarily close to the fixed point solution, then the semi-classical collapse phase (with the Hubble scale below the Planck scale) can last for many e-folds before diffusion drives the evolution away from the fixed point solution.


Our analysis is limited to first-order perturbations about the classical fixed points, however one might expect that fluctuations about the pressureless collapse eventually end up in a kinetic-energy-dominated collapse phase with $w=+1$ since this is the stable fixed point solution in this case. Perturbations about the kinetic-dominated collapse due to quantum fluctuations are adiabatic in the super-Hubble limit and so this fixed point solution remains stable against quantum noise up to the Planck scale.

In this paper, we focused on stochastic effects in the scalar field evolution in a collapsing cosmology.
Given that we have seen that quantum diffusion can play an important role in the semi-classical dynamics, it is interesting to consider whether quantum effects might be a way to avoid the classical singularity as the Hubble rate diverges. However it is not possible to model a cosmological bounce in our approach as we have used the scale factor as a monotonic time variable in order to study fluctuations in the scalar field. It might be interesting instead to use the scalar field (or some other quantity) as a monotonic time variable and then study fluctuations of the geometry, which might, in principle, include a non-monotonic scale factor, i.e., a bounce. We leave this open question for future work.

		
\subsection*{Acknowledgements}
	
The authors are grateful to Chris Pattison, Vincent Vennin, Misao Sasaki and Nelson Pinto-Neto for useful comments. This study was financed in part by the \emph{Coordena\c{c}\~ao de Aperfei\c{c}oamento de Pessoal de N\'ivel Superior} - Brazil (CAPES) - Finance Code 001. DW is supported in part by STFC grants ST/N000668/1 and ST/R505018/1.
	
\appendix

\section{Mapping between $p$, $\lambda^{2}$, $w$, $\nu$ and $n_{s}-1$}\label{AppMapping}
Throughout this work, we use the quantities $p$, $\lambda^{2}$, $w$ and $\nu$ because, even if they are connected to the others, each one of them is more appropriate for a specific analysis. In order to facilitate the understanding of the reader, we show the explicit mapping between them in 
Table \ref{table3}.
\begin{table}[h!]
	\centering
	\begin{tabular}{c|c|c|c|c|c}
		&\large $p$ \hspace{0.2cm}& \large $\lambda^{2}$ & \hspace{0.2cm} \large $w$ \hspace{0.2cm} & \large $\nu$ \hspace{0.2cm} & \large $n_{s}-1$ \\\hline\hline
		\large $p$ \hspace{0.2cm} & $p$ \hspace{0.2cm}& $\frac{2}{\lambda^{2}}$ \hspace{0.2cm}& $\frac{2}{3(1+w)}$ \hspace{0.2cm}& $\frac{2\nu -1}{2\nu - 3}$ \hspace{0.2cm}& $\frac{4-\left(n_{s}-1\right)}{6-\left(n_{s}-1\right)}$\\ \hline
		\large $\lambda^{2}$ \hspace{0.2cm} & $\frac{2}{p}$ \hspace{0.2cm}& $\lambda^{2}$ \hspace{0.2cm}& $3(1+w)$ \hspace{0.2cm}& $\frac{4\nu - 6}{2\nu -1}$ \hspace{0.2cm}& $\frac{2\left(n_{s}-1\right)-12}{\left(n_{s}-1\right)-4}$\\ \hline
		\large $w$ \hspace{0.2cm} & $\frac{2-3p}{3p}$ \hspace{0.2cm}& $\frac{\lambda^{2}-3}{3}$ \hspace{0.2cm}& $w$ \hspace{0.2cm}& $\frac{-2\nu - 3}{6\nu - 3}$ \hspace{0.2cm}& $\frac{\left(n_{s}-1\right)}{12-3\left(n_{s}-1\right)}$\\ \hline
		\large $\nu$ \hspace{0.2cm} & $\frac{3}{2}+\frac{1}{p-1}$ \hspace{0.2cm}& $\frac{3}{2}+\frac{\lambda^{2}}{2-\lambda^{2}}$ \hspace{0.2cm}& $\frac{3}{2}-\frac{3(1+w)}{1+w}$ \hspace{0.2cm} & $\nu$ \hspace{0.2cm}& $\frac{3\left(n_{s}-1\right)}{8}-\frac{3}{2}$ \\ \hline
		\large $n_{s}-1$ \hspace{0.2cm}& $\frac{6p-4}{p-1}$ \hspace{0.2cm}& $\frac{4(3-\lambda^2)}{2-\lambda^2}$ \hspace{0.2cm}& $\frac{12w}{1+3w}$ \hspace{0.2cm}& $\frac{4(2\nu+3)}{3}$ \hspace{0.2cm}& $n_{s}-1$
	\end{tabular}
	\caption{Table showing how to write $p$, $\lambda^{2}$, $w$, $\nu$ and $n_{s}-1$ in terms of each of them.}\label{table3}
\end{table}

\section{Kinetic-dominated solution}
\label{AppKinetic}

We show in this appendix the solutions for the other critical point, namely the kinetic-dominated regime. This is interesting for two reasons. First, this regime corresponds to the critical value $\nu=0$, which is the interface between purely adiabatic perturbations (at first-order) and non-adiabatic perturbations. Second, perturbations in this regime act as a stiff fluid and go as $a\propto (-\eta)^{-6}$. Such behaviour is usually invoked in the classical resolution of the initial singularity, see for instance the reviews \cite{battefeld2015critical,lilley2015bouncing,brandenberger2017bouncing,peter2008cosmology}. 

To begin, note we can rewrite (\ref{eq:delta x total}) in terms of the variables $x$ and $y$ as
\begin{align}
\delta x = \frac{\kappa}{6}\left[\left(1 - x^{2}\right)\frac{\dot{\delta\varphi}}{H} + \left(3x^{4}-3x^{2}+\frac{\lambda^{2}}{2}y^{2}\right)\delta\varphi\right]\;.
\end{align}
In the case of the kinetic-dominated solution, the fixed points are $x_a =\pm 1$, $y_a=0$. In this configuration, we have $w=1$, or equivalently $\lambda^2=6$, resulting in the trivial expression
\begin{align}
\delta x = 0 \;,
\end{align}
regardless of the value of the solution $\delta\varphi$. Hence, any first-order linear field perturbation leads to adiabatic perturbations in the kinetic-dominated regime.

\section{A different approach to find the noise}

\label{AppPerturbedMomentum}
\label{appendixc}	

We know that $\dot{\varphi}$ can be related to its momentum $\pi\equiv \partial {\cal L}/\partial \dot\varphi$ using the ADM formalism, as shown in \cite{Grain:2017dqa}, by
\begin{align}
\dot{\varphi} = \frac{1}{a(t)^{3}}\pi_{\varphi}\;,\label{eq:phipi2}
\end{align}
where the lapse function $N$ was chosen as the cosmic time, which means that $N=1$. Also, the evolution for $\pi_{\varphi}$ is given by
\begin{align}
\dot{\pi}_{\varphi} = - a(t)^{3} V_{,\varphi}
\;,
\end{align}
where ``,$\varphi$" represents the derivative with respect to $\varphi$.

The scalar field and its momentum can be split into a long-wavelength part and small-wavelength part as
\begin{align}
\varphi = \overline{\varphi} + \varphi_{Q} \quad \pi = \overline{\pi} + \pi_{Q}\;,
\end{align}
where the subscript ``$Q$" describes the small-wavelength part, which will allow us to calculate the quantum noise for $\varphi$ and $\pi$. 

We get the linearly perturbed momentum including scalar field perturbations
\begin{align}
\pi + \delta \pi = \frac{\partial \left( \mathcal{L}+\delta \mathcal{L}\right)}{ \partial \left(\frac{1}{1+A}\dot{\varphi} \right)} = \left(1+A\right) \frac{\partial \left( \mathcal{L}+\delta \mathcal{L}\right)}{ \partial \dot{\varphi}}
\end{align}
where we have also perturbed the lapse function $t\rightarrow(1+A)t$.
The perturbed momentum is then
\begin{align}
\delta \pi &= A \frac{\partial \mathcal{L}}{\partial \dot{\varphi}} + \frac{\partial \delta \mathcal{L}}{\partial \dot{\varphi}} \nonumber \\
&= a^3 \left(\dot{\delta \varphi}-A\dot{\varphi}\right)
\end{align}
We may use the constraint $A = \kappa^{2}\dot{\varphi}\delta\varphi/2H$ to eliminate the perturbed lapse function since we are working in the spatially-flat gauge \cite{Pattison:2019hef}.  

Using the definition of the coarse-graining scale \eqref{cgscale} in the expressions for the field and its conjugate momentum
\begin{align}
\delta\varphi &= \frac{i}{\sqrt{4\pi}}\left(\frac{2}{2\nu -1} \right) \frac{2^{|\nu|}\Gamma(\lvert\nu\rvert)}{k^{|\nu|}} \frac{H}{(-\eta)^{|\nu|-3/2}}\;,\label{eq:deltaphisol}\\
\delta\pi_{\varphi} &= \frac{i}{\sqrt{4\pi}}\frac{2^{|\nu|}\Gamma(\lvert\nu\rvert)}{k^{\lvert\nu\rvert}}\left(\nu - \frac{1}{2}\right)^{4}\left[\left(\frac{2}{2\nu - 1}\right)\left(\lvert\nu\rvert - \nu\right) -  \frac{\kappa^2 \dot{\varphi}^{2}}{2H^{2}}\right]\frac{1}{H(-\eta)^{\lvert\nu\rvert+3/2}} \;,\label{eq:pisol}
\end{align}
we get
\begin{align}
\lvert\delta\varphi\rvert^2 &= \frac{\Gamma^{2}(\lvert\nu\rvert)2^{2\lvert\nu\rvert}H^{2}}{4\pi \sigma^{2\lvert\nu\rvert}} \left(\frac{2}{2\nu - 1}\right)^{2\lvert\nu\rvert + 2}\frac{1}{(-\eta)^{-3}}\;, \\
\lvert\delta\pi\rvert^2 &= \frac{\Gamma^{2}(\lvert\nu\rvert)2^{2\lvert\nu\rvert}}{4\pi \sigma^{2\lvert\nu\rvert} H^{2}}\left(\frac{2}{2\nu - 1}\right)^{2\lvert\nu\rvert -4}\left[\left(\frac{2}{2\nu - 1}\right)\left(\lvert\nu\rvert - \nu\right) - 4\pi G \frac{\dot{\varphi}^{2}}{H^{2}}\right]^{2}\frac{1}{(-\eta)^{3}}\;, \\
\delta\varphi \delta\pi^{*} & = \frac{\Gamma^{2}(\lvert\nu\rvert)2^{2\lvert\nu\rvert}}{4\pi \sigma^{2\lvert\nu\rvert}} \left(\frac{2}{2\nu - 1}\right)^{2\lvert\nu\rvert -1} \left[\left(\frac{2}{2\nu - 1}\right)\left(\lvert\nu\rvert - \nu\right) -4\pi G \frac{\dot{\varphi}^{2}}{H^{2}}\right]\;.
\end{align}

From this we can work out the two-points correlation matrix associated with the quantum noise with respect to conformal time $\xi_{\varphi}$ and $\xi_{\pi}$ \cite{Grain:2017dqa}
.\begin{align}
\Xi_{\varphi, \varphi}^{(\eta)} =& \frac{1}{(2\pi)^{3}}\frac{\Gamma^{2}(\lvert\nu\rvert)2^{2\lvert\nu\rvert}}{\sigma^{2\lvert\nu\rvert - 3}} \left(\frac{2}{2\nu -1}\right)^{2\lvert\nu\rvert -1}\frac{H^{2}(\eta)}{(-\eta)} \;,\\
\Xi_{\pi, \pi}^{(\eta)} =& \frac{1}{(2\pi)^{3}}\frac{\Gamma^{2}(\lvert\nu\rvert)2^{2\lvert\nu\rvert}}{\sigma^{2\lvert\nu\rvert - 3}}\left(\frac{2}{2\nu -1}\right)^{2\lvert\nu\rvert -7}\nonumber\\
&\times \left[\left(\frac{2}{2\nu - 1}\right)\left(\lvert\nu\rvert - \nu\right) -  4\pi G \frac{\dot{\varphi}^{2}}{H^{2}(\eta)}\right]^{2}\frac{1}{H^{2}(\eta)(-\eta)^{7}} \;,\\
\Xi_{\varphi, \pi}^{(\eta)} = \Xi_{\pi, \varphi}^{(\eta)} =& \frac{1}{(2\pi)^{3}}\frac{\Gamma^{2}(\lvert\nu\rvert)2^{2\lvert\nu\rvert}}{\sigma^{2\lvert\nu\rvert - 3}} \left(\frac{2}{2\nu -1}\right)^{2\lvert\nu\rvert -4}\nonumber\\
&\times \left[\left(\frac{2}{2\nu - 1}\right)\left(\lvert\nu\rvert - \nu\right) -  4\pi G \frac{\dot{\varphi}^{2}}{H^{2}(\eta)}\right]\frac{1}{(-\eta)^{4}}\;,
\end{align}
with the dependence in conformal time is left explicityly in the subscript. We can write a stochastic version for $x$ of the form
\begin{align}
	\dot{x} = \bar{x}+\xi_x \;,
\end{align}
and relate the noise of $x$ to the noises of $\varphi$ and $\pi_{\varphi}$. By doing so, it can be shown the correlation matrix in $x$ is a combination of those contributions,
\begin{align}
\label{xix}
\left\langle \xi_{x}  \xi_{x} \right\rangle := \Xi_{x,x}^{(\eta)} = \frac{a^{12}}{(\bar{\pi}_{\varphi}^2+2a^6V)^{3}} \left[4V^2 \Xi_{\pi, \pi}^{(\eta)}+(\bar{\pi}_{\varphi}V^{\prime})^2\Xi_{\varphi, \varphi}^{(\eta)} -4VV^{\prime}\bar{\pi}_{\varphi} \Xi_{\varphi, \pi}^{(\eta)}\right] \;,
\end{align}
and using the expression of the correlation matrix \eqref{eq:Xifg} in terms of number of e-folds as
\begin{align}
\Xi_{x,x}^{(N)}&=\frac{d\eta}{dN} \; \Xi_{x,x}^{(\eta)} \nonumber\\
&= (-\eta_{\star}) \frac{1}{\nu-1/2} \exp\left[\frac{1}{\nu-1/2}\left(N_{\star}-N\right)\right]  \Xi_{x,x}^{(\eta)}\;,
\end{align}
it is straigthforward to show we can recover our result \eqref{deltaxN}.



\section{Next-to-leading order field contribution}
\label{appnextorder}
We expand the field solution \eqref{eq:delta_phi} and its derivative to third order to get all terms contributing to second order. Then the field is now
\begin{align}
\delta \varphi &= \frac{i}{a} \frac{\Gamma\left(|\nu|\right)2^{|\nu|}}{\sqrt{4\pi}k^{|\nu|}} \left[1+\frac{\left(-k\eta\right)^2}{4\left(|\nu|-1\right)}+\frac{\left(-k\eta\right)^4}{32\left(|\nu|-1\right)\left(|\nu|-2\right)}\right] \frac{1}{\left(-\eta\right)^{|\nu|-1/2}} \nonumber \\
&= \frac{i}{\sqrt{4\pi}}\left(\frac{2}{2\nu -1} \right) \frac{2^{|\nu|}\Gamma(\lvert\nu\rvert)}{k^{|\nu|}} \nonumber \\
&\quad \times \left[ \frac{H}{(-\eta)^{|\nu|-3/2}}+ \frac{k^2H}{4\left(|\nu|-1\right) (-\eta)^{|\nu|-7/2}} + \frac{k^4H}{32\left(|\nu|-1\right)\left(|\nu|-2\right) (-\eta)^{|\nu|-11/2}}\right] \;, \\
\dot{\delta \varphi} &= \frac{i}{\sqrt{4\pi}}\left(\frac{2}{2\nu -1} \right)^2 \frac{2^{|\nu|}\Gamma(\lvert\nu\rvert)}{k^{|\nu|}} H^2 \nonumber \\
&\quad \times \left[\frac{\left(|\nu|-\nu\right)}{ (-\eta)^{|\nu|-3/2}}+ \frac{k^2 \left(|\nu|-\nu-2\right)}{4\left(|\nu|-1\right) (-\eta)^{|\nu|-7/2}} + + \frac{k^4 \left(|\nu|-\nu-4\right)}{32\left(|\nu|-1\right)\left(|\nu|-2\right) (-\eta)^{|\nu|-11/2}}\right] \;.
\end{align}

The contribution to the noise in $x$ becomes
\begin{align}
\Xi_{x,x}(N) &= \bar{g}(\nu,\sigma) \kappa^{2} H^{2}_{\star} \exp{\left[-\frac{3-2\nu}{\nu-1/2} \left(N_{\star}-N\right)\right]}
\;,
\end{align}
with
\begin{align}
\bar{g}(\nu,\sigma) :=  &\frac{\Gamma^{2}(\lvert\nu\rvert)2^{2\lvert\nu\rvert +2}}{(12\pi)^{3}}\left(\frac{\nu^{2}}{\sigma^{2\lvert\nu\rvert - 3}}\right)\left(\frac{2}{2\nu - 1}\right)^{2\lvert\nu\rvert+4} \left[\left(\lvert\nu\rvert - \nu\right)^{2} +\frac{\sigma^2}{2} \frac{\left(\lvert\nu\rvert - \nu\right)\left(\lvert\nu\rvert - \nu -2\right)}{\left(\lvert\nu\rvert - 1\right)} \left(\nu - \frac{1}{2}\right)^2 \right. \nonumber \\
&\left. +\frac{\sigma^4}{16} \frac{\left[\left(\lvert\nu\rvert - \nu\right)^2-4\left(\lvert\nu\rvert - \nu\right)\right]\left[2\lvert\nu\rvert-3\right]+4\left(\lvert\nu\rvert-2\right)}{\left(\lvert\nu\rvert - 1\right)^2\left(\lvert\nu\rvert - 2\right)} \left(\nu - \frac{1}{2}\right)^4 \right]\;.
\end{align}
For positive $\nu$, the above equation is simplified to
\begin{align}
\bar{g}(\nu > 0,\sigma) :=  &\frac{\Gamma^{2}(\lvert\nu\rvert)2^{2\lvert\nu\rvert +2}}{(12\pi)^{3}\sigma^{2\lvert\nu\rvert -7}}\left(\frac{\nu}{\lvert\nu\rvert - 1}\right)^{2}\left(\frac{2}{2\nu -1}\right)^{2\lvert\nu\rvert} \;.
\end{align}
The variance in the case of quasi-de Sitter inflation  ($\nu=3/2-\epsilon$) is given by
\begin{align}
\sigma_{x,qu}^2 &= \frac{1}{24 \pi^2}  H_{\star}^2 \kappa^2 \sigma^{4-\epsilon}\left(1+4\epsilon\right) \left\lbrace 1- \exp{\left[-6(N-N_{\star})\right]}\right\rbrace \;.
\end{align}
Since $N$ is growing with time the exponential vanishes quickly and eventually the time at which $\sigma_{x,qu}^2=1$ is 
\begin{align}
H(N)=\sqrt{3\pi}\left(1-2\epsilon\right)\frac{M_{pl}}{\sigma^{2-\epsilon/2}} \;,
\end{align}
and since $\sigma<1$ we see quantum diffusion drives us away only if the Hubble rate is far above the Planck scale. We note this result stays true for $N\approx N_{\star}$ since in this case we have
\begin{align}
H(N)=\sqrt{\frac{\pi}{2}}\left(1-2\epsilon\right)\frac{M_{pl}}{\sigma^{2-\epsilon/2}} \;.
\end{align}
	
\section{Fourier transform on a finite domain}
\label{App2}
This Appendix is dedicated to show explicitly the solution given by (\ref{variance}), while pointing out that the final result depends on a conventional factor. From (\ref{variancedifeq}), we easily get 
\begin{align}
\label{stochastic variance}
\sigma_x^2 (N) = \sigma_x^2(N_{\star}) e^{2m\left(N-N_{\star}\right)} + 2\int_{N_{\star}}^{N}  dS \, e^{2m(N-S)} \left\langle \hat{\xi}_x(S) \left(\bar{x}(S)-x_c\right) \right\rangle.
\end{align}
We can reexpress the expectation value in the second term using \eqref{stochastic solution x} to get
\begin{align}
\left\langle \hat{\xi}_x(S) \left(\bar{x}(S)-x_c\right) \right\rangle = \left\langle\int_{S_{\star}}^{S} e^{m\left(S-U\right)} \hat{\xi}_x(S)\hat{\xi}_{x}(U) dU \right\rangle\;.
\end{align}
Using the Fubini theorem, we get
\begin{align}
\left\langle \hat{\xi}_x(S) \left(\bar{x}(S)-x_c\right) \right\rangle &= \int_{S_{\star}}^{S} e^{m\left(S-U\right)} \left\langle \hat{\xi}_x(S)\hat{\xi}_{x}(U)  \right\rangle dU \;.
\end{align}
The variance \eqref{stochastic variance} is then
\begin{align}
\sigma_x^2 (N) = \sigma_x^2(N_{\star}) e^{2m\left(N-N_{\star}\right)} + 2\int_{N_{\star}}^{N} \int_{S_{\star}}^{S} dS dU \, e^{m(2N-S-U)} \left\langle \hat{\xi}_x(S)\hat{\xi}_{x}(U) \right\rangle \;.
\end{align}
The resolution of \eqref{stochastic variance} leads us to consider the following integral:
\begin{align}
\int_{S_{\star}}^{S} dU e^{m\left(2N-S-U\right)} \delta 
\left(S-U\right) \;.
\end{align}
For a general function, we have
\begin{align}
\int_{a}^{b} f(x) \delta (x-x^{\prime}) dx &= \int_{-\infty}^{\infty} dx \, f(x) \left[\theta(x-a)-\theta(x-b)\right]\delta (x-x^{\prime})  \nonumber \\
&= f(x^{\prime}) \left[\theta(x^{\prime}-a) -\theta(x^{\prime}-b) \right] \;,
\end{align}
what gives in our case
\begin{align}
\int_{S_{\star}}^{S} dU e^{m\left(2N-S-U\right)} \delta 
\left(S-U\right) &= e^{2m\left(N-S\right)} \left[\theta(S-S_{\star}) -\theta(S-S)\right] \nonumber \\
&= e^{2m\left(N-S\right)} \left[\theta(S-S_{\star}) -\theta(0)\right] \;. 
\end{align}
Using the half-maximum convention for the unit step function, we obtain
\begin{align}
\int_{S_{\star}}^{S} dU e^{m\left(2N-S-U\right)} \delta 
\left(S-U\right) &= \frac{1}{2} e^{2m\left(N-S\right)} \;.
\end{align}
Now, we are able to write the full solution for the variance in $x$ as
\begin{align}
\sigma_x^2 (N) = \sigma_x^2(N_{\star}) e^{2m\left(N-N_{\star}\right)} + \int_{N_{\star}}^{N}  dS \, e^{2m(N-S)} \Xi_{x,x}(S)  \;.
\end{align}

\bibliographystyle{unsrturl}
\bibliography{stochasticcollapse}

\end{document}